# Network Slicing for 5G with SDN/NFV: Concepts, Architectures and Challenges

J. Ordonez-Lucena[1], P. Ameigeiras[1,2], D. Lopez[3], J.J. Ramos-Munoz[1,2], J. Lorca[4], J. Folgueira[5]

[1]*Department of Signal Theory, Telematics, and Communications, University of Granada.*
[2]*Research Centre for Information and Communications Technologies, University of Granada.*
[3]*Technology Exploration & Standards, Telefónica I+D – Global CTO.*
[4]*Radio Access Networks–Innovation, Telefónica I+D – Global CTO.*
[5]*Transport & IP Networks, Telefónica I+D – Global CTO.*

*Abstract*— The fifth generation of mobile communications is anticipated to open up innovation opportunities for new industries such as vertical markets. However, these verticals originate myriad use cases with diverging requirements that future 5G networks have to efficiently support. Network slicing may be a natural solution to simultaneously accommodate over a common network infrastructure the wide range of services that vertical-specific use cases will demand. In this article, we present the network slicing concept, with a particular focus on its application to 5G systems. We start by summarizing the key aspects that enable the realization of so-called network slices. Then, we give a brief overview on the SDN architecture proposed by the ONF and show that it provides tools to support slicing. We argue that although such architecture paves the way for network slicing implementation, it lacks some essential capabilities that can be supplied by NFV. Hence, we analyze a proposal from the ETSI to incorporate the capabilities of SDN into the NFV architecture. Additionally, we present an example scenario that combines SDN and NFV technologies to address the realization of network slices. Finally, we summarize the open research issues with the purpose of motivating new advances in this field.

*Keywords – 5G, Network Slicing, SDN, NFV*

## 1. Introduction

5G systems are nowadays being investigated to satisfy the consumer, service and business demands of 2020 and beyond. One of the key drivers of 5G systems is the need to support a variety of vertical industries such as manufacturing, automotive, healthcare, energy, and media & entertainment [1]. Such verticals originate very different use cases, which impose a much wider range of requirements than existing services do nowadays. Today's networks, with their "one-size-fits-all" architectural approach, are unable to address the diverging performance requirements that verticals impose in terms of latency, scalability, availability and reliability. To efficiently accommodate vertical-specific use cases along with increased demands for existing services over the same network infrastructure, it is accepted that 5G systems will require architectural enhancements with respect to current deployments.

Network softwarization, an emerging trend which seeks to transform the networks using software-based solutions, can be a potential enabler for accomplishing this. Through technologies like Software-Defined Networking (SDN) and Network Function Virtualization (NFV), network softwarization can provide the programmability, flexibility, and modularity that is required to create multiple logical (virtual) networks, each tailored for a given use case, on top of a common network. These logical networks are referred to as *network slices*. The concept of separated virtual networks deployed over a single network is indeed not new (e.g. VPN), although there are specificities that make network slices a novel concept. We define network slices as end-to-end (E2E) logical networks running on a common underlying (physical or virtual) network, mutually isolated, with independent control and management, and which can be created on demand. Such self-contained networks must be flexible enough to simultaneously accommodate diverse business-driven use cases from multiple players on a common network infrastructure (see Figure 1).

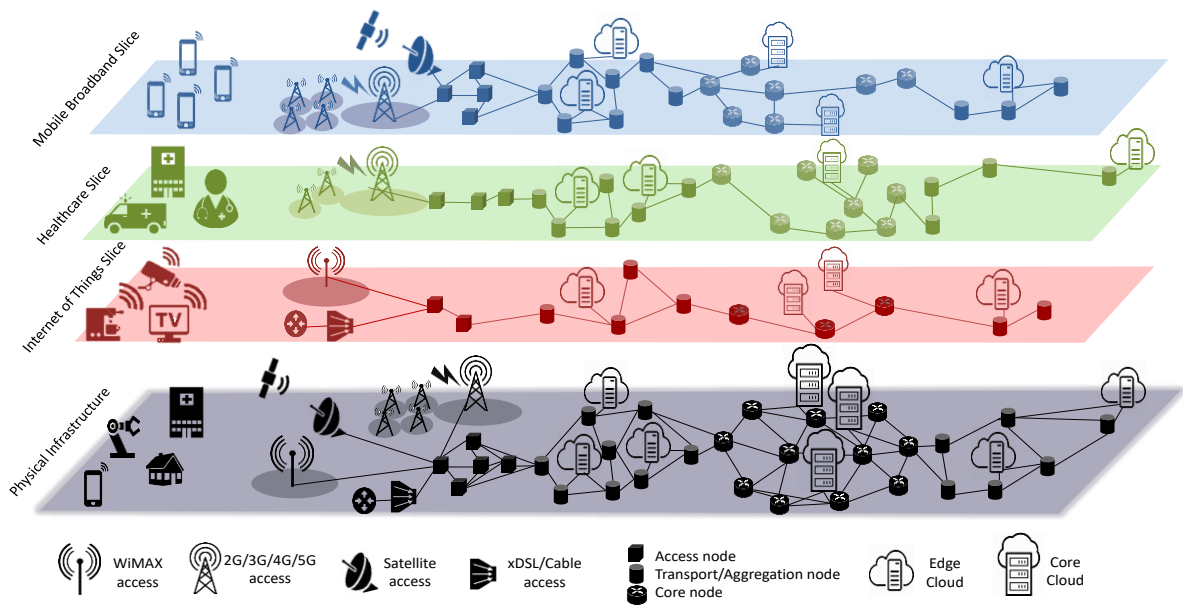

Figure 1. 5G network slices running on a common underlying multi-vendor and multi-access network. Each slice is independently managed and addresses a particular use case.

In this paper, we provide a comprehensive study of the architectural frameworks of both SDN and NFV as key enablers to achieve the realization of network slices. Although these two approaches are not yet commonplace in current networking practice, especially in public wide area networks (WANs), their integration offers promising possibilities to adequately meet the slicing requirements. Indeed, many 5G research and demonstration projects (such as 5GNORMA, 5GEx, 5GinFIRE, or 5G!Pagoda) are addressing the realization of 5G slicing through the combination of SDN and NFV. Thus, we present a deployment example that illustrates how NFV functional blocks, SDN controllers, and their interactions can fully realize the network slicing concept. Furthermore, we identify the main challenges arising from implementing network slicing for 5G systems.

The remainder of this paper is organized as follows: Section 2 provides a background on key concepts for network slicing. Sections 3 and 4 describe the SDN architecture from the ONF and the NFV architecture from the ETSI, respectively. Section 5 shows a network slicing use case with NFV and SDN integration, and Section 6 provides the main challenges and future research directions.

## 2. Background on key concepts for Network Slicing

In this section, we provide a background on key aspects that are necessary to realize the network slicing concept.

### 2.1 Resources

In its general sense, a resource is a manageable unit, defined by a set of attributes or capabilities that can be used to deliver a service. A network slice is composed of a collection of resources that, appropriately combined together, meet the service requirements of the use case that such slice supports. In network slicing, we consider two types of resources:

- Network Functions (NFs): functional blocks that provide specific network capabilities to support and realize the particular service(s) each use case demands. Generally implemented as software instances running on infrastructure resources, NFs can be physical (a combination of vendor-specific hardware and software, defining a traditional purpose-built physical appliance) and/or virtualized (network function software is decoupled from the hardware it runs on).
- Infrastructure Resources: heterogeneous hardware and necessary software for hosting and connecting NFs. They include computing hardware, storage capacity, networking resources (e.g. links and switching/routing devices enabling network connectivity) and physical assets for radio access. Suitable for being used in network slicing, the aforementioned resources and their attributes have to be abstracted and logically partitioned leveraging virtualization mechanisms, defining virtual resources that can be used in the same way as physical ones.

## 2.2 Virtualization

Virtualization is a key process for network slicing as it enables effective resource sharing among slices. Virtualization is the abstraction of resources using appropriate techniques. Resource abstraction is the representation of a resource in terms of attributes that match predefined selection criteria while hiding or ignoring aspects that are irrelevant to such criteria, in an attempt to simplify the use and management of that resource in some useful way. The resources to be virtualized can be physical or already virtualized, supporting a recursive pattern with different abstraction layers.

Just as server virtualization [2] makes virtual machines (VMs) independent of the underlying physical hardware, network virtualization [3] enables the creation of multiple isolated virtual networks that are completely decoupled from the underlying physical network, and can safely run on top of it.

The introduction of virtualization to the networking field enables new business models, with novel actors and distinct business roles. We consider a framework with three kinds of actors:

- Infrastructure Provider (InP): owns and manages a given physical network and its constituent resources. Such resources, in form of WANs and/or data centers (DCs), are virtualized and then offered through programming interfaces to a single or multiple tenants.
- Tenant: leases virtual resources from one or more InPs in the form of a virtual network, where the tenant can realize, manage and provide network services to its users. A network service is a composition of NFs, and it is defined in terms of the individual NFs and the mechanism used to connect them.
- End user: consumes (part of) the services supplied by the tenant, without providing them to other business actors.

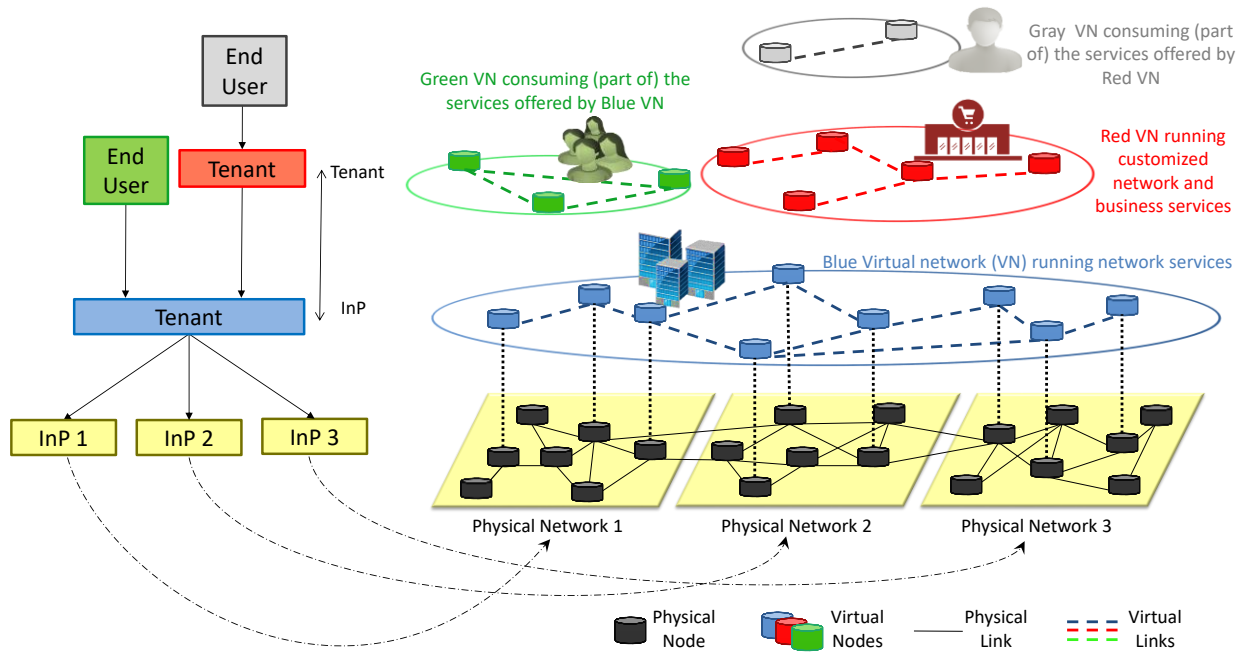

Figure 2. InPs and tenants as virtualization actors. This scenario shows the recursion principle, where these actors happen in a vertical multi-layered pattern.

As discussed above, virtualization is naturally recursive, and the first two actors can happen in a vertical multi-layered pattern, where a tenant at one layer acts as the InP at the layer immediately above. The recursion mentioned here implies that a tenant can provide network services to an end user, but also to another tenant (see Figure 2). In such a case, this second tenant would provide more advanced network services to its own users.

## 2.3 Orchestration

Orchestration is also a key process for network slicing. In its general sense, orchestration can be defined as the art of both bringing together and coordinating disparate things into a coherent whole. In a slicing environment, where the players involved are so diverse, an orchestrator is needed to coordinate seemingly disparate network processes for creating, managing and delivering services.

A unified vision and scope of orchestration has not been agreed upon. According to the Open Network Foundation (ONF) [4], orchestration is defined as the continuing process of selecting resources to fulfill client service demands in an optimal manner. The idea of optimal refers to the optimization policy that governs orchestrator behavior, which is expected to meet all the specific policies and SLAs associated with clients (e.g. tenants or end users) that request services. The term continuing means that available resources, service demands and optimization criteria may change in time. Interestingly, orchestration is also referred in [4] as the defining characteristic of an SDN controller. Note that client is a term used in SDN context.

The ONF states that the orchestrator functions include client-specific service demand validation, resource configuration, and event notification. For a more detailed description of these functions, see Section 6.2 in [4].

However, in network slicing orchestration cannot be performed by a single centralized entity, not only because of the complexity and broad scope or orchestration tasks, but also because it is necessary to preserve management independence and support the possibility of recursion. In our view, a framework in which each virtualization actor (see Section 2.2) has an entity performing orchestration functions seems more suitable to satisfy the above requirements. The entities should exchange information and delegate functionalities between them to ensure that the services delivered at a certain abstraction layer satisfy the required performance levels with optimal resource utilization.

## 2.4 Isolation

Strong isolation is a major requirement that must be satisfied to operate parallel slices on a common shared underlying substrate. The isolation must be understood in terms of:
- Performance: each slice is defined to meet particular service requirements, usually expressed in the form of KPIs. Performance isolation is an E2E issue, and has to ensure that service-specific performance requirements are always met on each slice, regardless of the congestion and performance levels of other slices.
- Security and privacy: attacks or faults occurring in one slice must not have an impact on other slices. Moreover, each slice must have independent security functions that prevent unauthorized entities to have read or write access to slice-specific configuration/management/accounting information, and able to record any of these attempts, whether authorized or not.
- Management: each slice must be independently managed as a separate network.

To achieve isolation, a set of appropriate, consistent policies and mechanisms have to be defined at each virtualization level, following the ideas introduced in Section 2.3. The policies (what is to be done) contain lists of rules that describe how different manageable entities must be properly isolated, without delving into how this can be achieved. The mechanisms (how it is to be done) are the processes that are implemented to enforce the defined policies. From our point of view, to fully realize the required isolation level, the interplay of both virtualization and orchestration is needed.

# 3. ONF Network Slicing Architecture

The SDN architecture provided by the ONF comprises an intermediate control plane that dynamically configures and abstracts the underlying forwarding plane resources so as to deliver tailored services to clients located in the application plane (see SDN basic model in [4]). This is well aligned with the requirements of 5G network slicing, which needs to satisfy a wide range of service demands in an agile and cost-effective manner. Thus, the SDN architecture is an appropriate tool for supporting the key principles of slicing. The purpose of this section is to describe the SDN architecture and how it can be applied to enable slicing in 5G systems.

According to [4], the major SDN architectural components are resources and controllers. For SDN, a resource is anything that can be utilized to provide services in response to client requests. This includes infrastructure resources and NFs (see Section 2.1), but also network services, in application of the recursion principle described in Section 2. A controller is a logically centralized entity instantiated in the control plane which run-time operates SDN resources to deliver services in an optimal way. Therefore, it mediates between clients and resources, acting simultaneously as server and client via client and server contexts, respectively. Both contexts are conceptual components of an SDN controller enabling the server-client relationships (see Figure 3):

- Client context: represents all the information the controller needs to support and communicate with a given client. It comprises a Resource Group and a Client support function. The Resource Group contains an abstract, customized view of all the resources that the controller, through one of its northbound interfaces, offers to the client, in order to deliver on its service demands and facilitate its interaction with the controller. Client support contains all that is necessary to support client operations, including policies on what the client is allowed to see and do [4], and service-related information to map actions between the client and the controller.
- Server context: represents all the information the controller needs to interact with a set of underlying resources, assembled in a Resource Group, through one of its southbound interfaces.

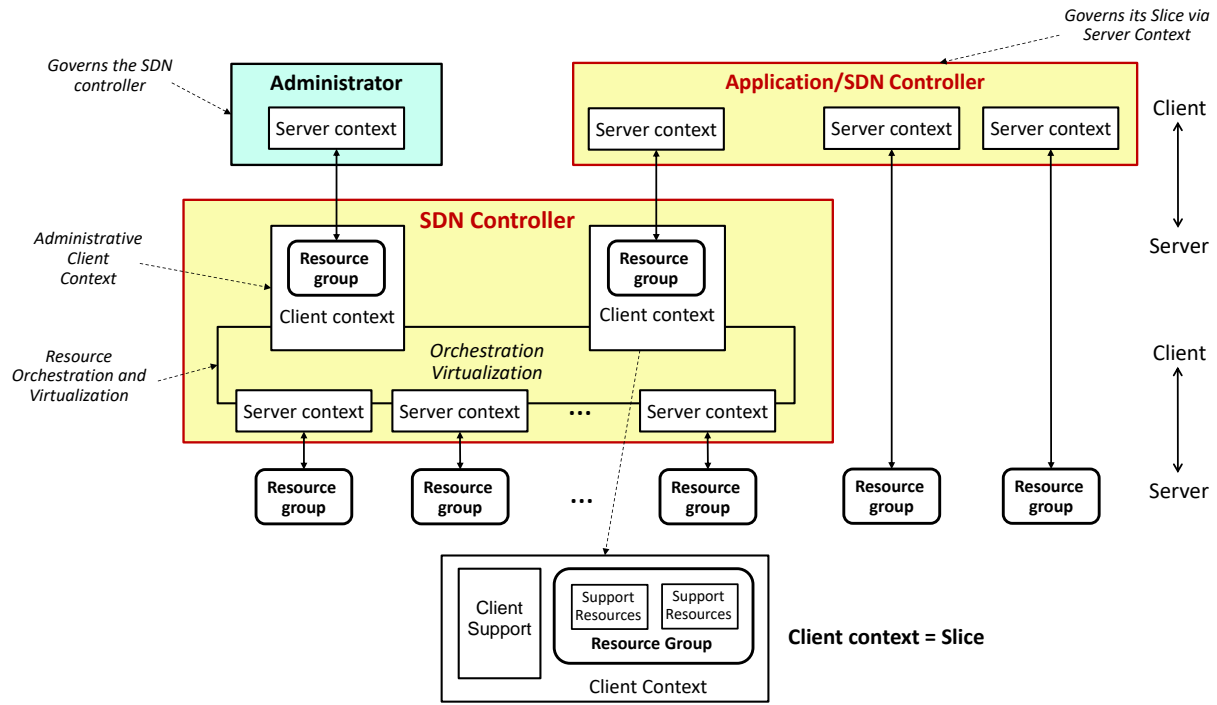

Figure 3. ONF SDN Network Slicing architecture [5].

The process of transforming the set of Resource groups accessed through server contexts to those defined in separate client contexts is not straightforward, and it requires the SDN controller to perform virtualization and orchestration functions.

When performing the virtualization function, the SDN controller carries out the abstraction and the aggregation/partitioning of the underlying resources. Thanks to virtualization, each client context provides a specific Resource Group that can be used by the client associated with that context to realize its service(s). Through the orchestration (see Section 2.3), the SDN controller optimally dispatches the selected resources to such separate Resource Groups. The interplay of both controller functions enables the fulfillment of the diverging service demands from all clients while preserving the isolation among them.

The SDN architecture also includes an administrator. Its tasks consist of instantiating and configuring the entire controller, including the creation of both server and client contexts and the installation of their associated policies.

According to the ONF vision, the SDN architecture naturally supports slicing [5], as the client context provides the complete abstract set of resources (as Resource Group) and supporting control logic that constitutes a slice, including the complete collection of related client service attributes.

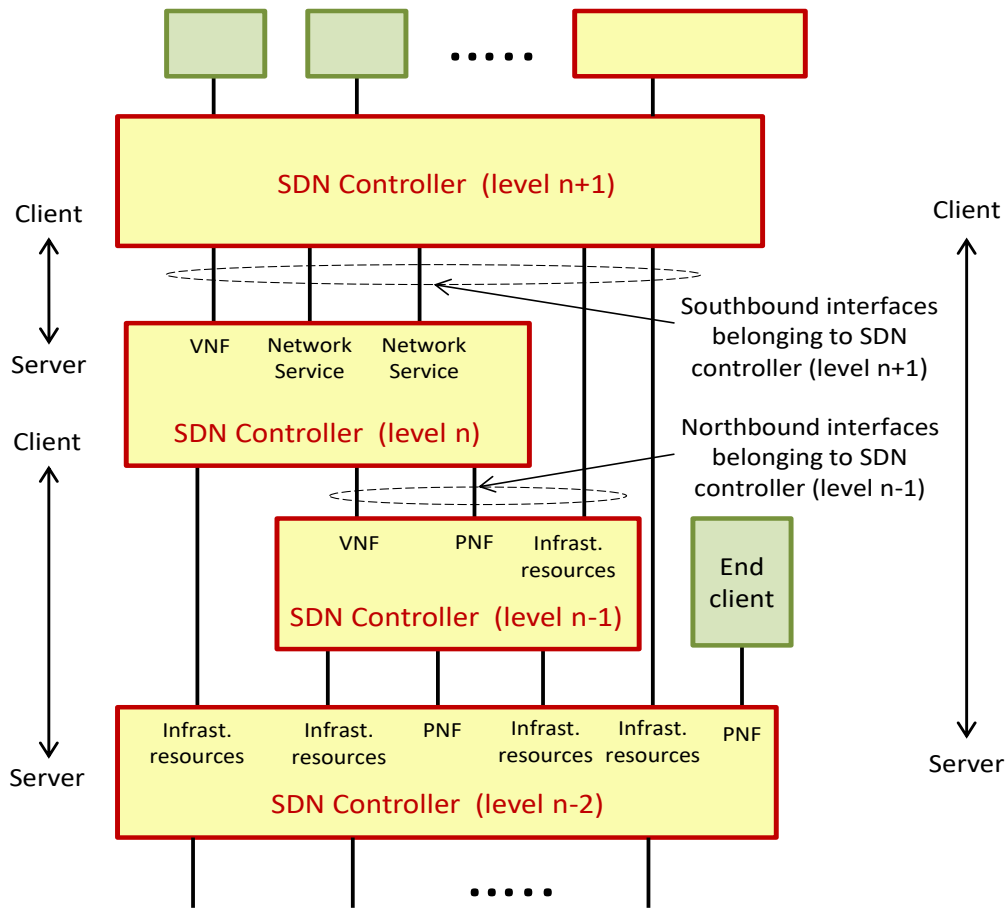

Figure 4. Complex client-server relationships enabled by the recursion in the SDN control plane, adapted from [7].

Another key functional aspect that makes SDN architecture ideal to embrace 5G slicing is recursion. Because of the different abstraction layers that the recursion principle enables, the SDN control plane can involve multiple hierarchically arranged controllers that extend the client-server relationships at several levels (see Figure 4). According to these premises, it is evident that SDN can support a recursive composition of slices [5]. This implies that the resources (i.e. Resource Group) a given controller delivers to one of its clients in form of a dedicated slice (i.e. client context) can, in turn, be virtualized and orchestrated by such client in case of being an SDN controller. This way, the new controller can utilize the resource(s) it access via its server context(s) to define, scale and deliver new resources (and hence new slices) to its own clients, which might also be SDN controllers.

# 4. NFV Reference Architectural framework

Although the SDN architecture described in Section 3 gives a comprehensive view of the control plane functionalities enabling slicing, it lacks capabilities that are vital to efficiently manage the lifecycle of network slices and its constituent resources. In this respect, the NFV architecture [6] is ideal to play this role, as it manages the infrastructure resources and orchestrates the allocation of such resources needed to realize VNFs and network services.

To take benefit from the management and orchestration functionalities from NFV, an appropriate cooperation between SDN and NFV is required. However, embracing SDN and NFV architectures into a common reference framework is not an easy task [7]-[8]. In this section, we briefly describe the tentative framework that ETSI presents in [8] to integrate SDN within the reference NFV architecture. This framework incorporates two SDN controllers, one logically placed at the tenant and another at the InP level. We commence providing a brief overview of the NFV architectural framework, and later describe the integration of the two SDN controllers (see Figure 5).

The NFV architecture comprises the following entities:

- Network Functions Virtualization Infrastructure (NFVI): a collection of resources used to host and connect the VNFs. While the broad scope of SDN makes resource a generic concept (see Section 3), the current resource definition in the NFV framework comprises only the infrastructure resources.
- VNFs: software-based implementation of NFs which run over the NFVI.
- Management and Orchestration (MANO): performs all the virtualization-specific management, coordination and automation tasks in the NFV architecture. The MANO framework [9] comprises three functional blocks:
  - Virtualized Infrastructure Manager (VIM): responsible for controlling and managing the NFVI resources.
  - VNF Manager (VNFM): performs configuration and lifecycle management of the VNF(s) on its domain.
  - Orchestrator: according to ETSI, it has two set of functions performed by Resource Orchestrator (RO) and Network Service Orchestrator (NSO) respectively. RO orchestrates the NFVI resources across (potentially different) VIMs. NSO performs the lifecycle management of network services, using the capabilities provided by the RO and the (potentially different) VNFMs.
- Network Management System (NMS): framework performing the general network management tasks. Although its functions are orthogonal to those defined in MANO, NMS

is expected to interact with MANO entities by means of a clear separation of roles [9]. NMS comprises:
- Element Management (EM): anchor point responsible for the FCAPS (Fault, Configuration, Accounting, Performance, and Security) of a VNF.
- Operation/Business Support System (OSS/BSS): a collection of systems and management applications that network service providers use to provision and operate their network services. In terms of the roles we consider in Section 2, tenants would run these applications.

ETSI proposal includes two SDN controllers in the architecture. Each controller centralizes the control plane functionalities and provides an abstract view of all the connectivity-related components it manages. These controllers are:

- Infrastructure SDN controller (IC): it sets up and manages the underlying networking resources to provide the required connectivity for communicating the VNFs (and its components [10]). Managed by the VIM, this controller may change infrastructure behavior on-demand according to VIM specifications, adapted from tenant requests.
- Tenant SDN controller (TC): instantiated in the tenant domain [11] as one of the VNFs or as part of the NMS, this second controller dynamically manages the pertinent VNFs used to realize the tenant's network service(s). These VNFs are the underlying forwarding plane resources of the TC. The operation and management tasks that the TC carries out are triggered by the applications running on top of it, e.g. the OSS.

Both controllers manage and control their underlying resources via programmable southbound interfaces, implementing protocols like OpenFlow, NETCONF or I2RS. However, each controller provides a different level of abstraction. While the IC provides an underlay to support the deployment and connectivity of VNFs, the TC provides an overlay comprising tenant VNFs that, properly composed, define the network service(s) such tenant independently manages on its slice(s). These different resource views each controller offers through its interfaces have repercussions on the way they operate. On one side, the IC is not aware of the number of slices that utilize the VNFs it connects, nor the tenant(s) which operates such slices. On the other side, for the TC the network is abstracted in terms of VNFs, without notions of how those VNFs are physically deployed. Despite their different abstraction levels, both controllers have to coordinate and synchronize their actions [8]. Note that the service and tenant concept mentioned here can be extended to higher abstraction layers by simply applying the recursion principle, as shown in Figure 2.

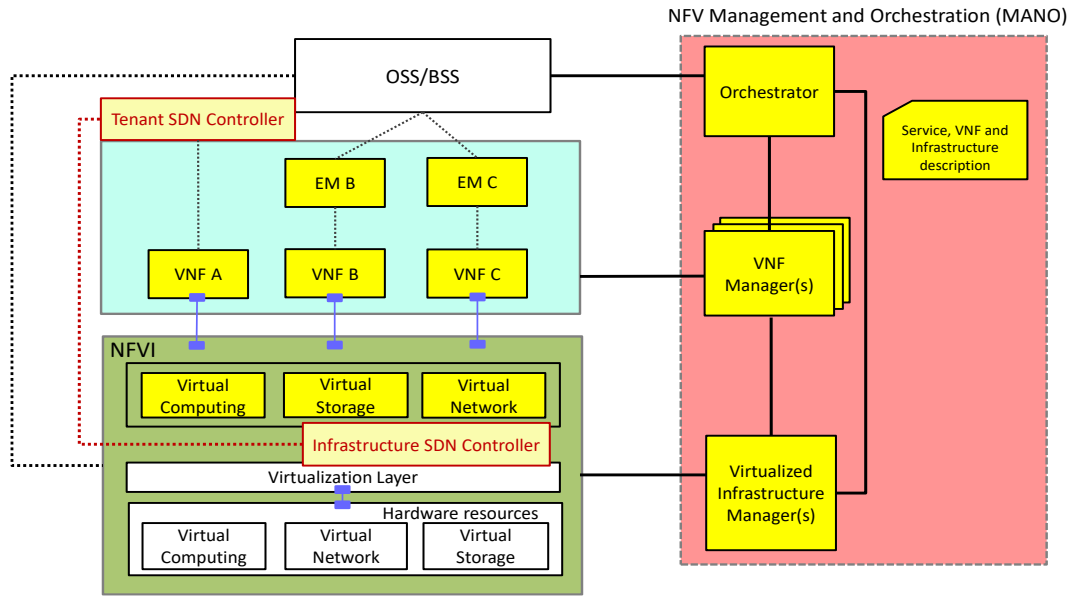

Figure 5. Integrating SDN controllers into the reference NFV architectural framework at the two levels required to achieve slicing.

## 5. Network Slicing use case with SDN-NFV Integration

In this section, we describe an SDN-enabled NFV deployment example that illustrates the network slicing concept, with several slices running on a common NFVI (see Figure 6). This deployment includes two tenants, each managing a particular set of slices. In the example, we only consider a single level of recursion, and thus the tenants directly serve the end users. Each slice consists of VNFs that are appropriately composed and chained to support and build up the network service(s) the slice (and thus the tenant) delivers to its users. Note that the deployment includes two distinct phases. First, a slice creation phase, in which an end user requests a slice from a network slice catalog, and then the tenant instantiates the slice. Next, a run-time phase, where the different functional blocks within each slice have already been created and are now operative. For simplicity, in Figure 6 we only depict the run-time phase.

The example considers that the tenants access NFVI resources from three InPs. InP1 provides compute and networking resources, both deployed on two NFVI-Points of Presence (NFVI-PoPs) [12] in the form of DCs. InP2 and InP3 provide SDN-based WAN transport networks, used to communicate such NFVI-PoPs. The VMs and their underlying hardware, instantiated in the NFVI-PoPs and in charge of hosting VNFs (and their components), are directly managed by the VIMs. The networking resources, supporting VM (and hence VNF) connectivity at the infrastructure level, are programmatically managed by the ICs following the VIM and the WAN infrastructure manager (WIM) premises. Both VIMs and WIMs act as SDN applications, delegating the tasks related to the management of networking resources to their underlying ICs. Although in this example the ICs are deployed on the NFVI, it would be possible to integrate them into their corresponding VIMs, as [8] suggests.

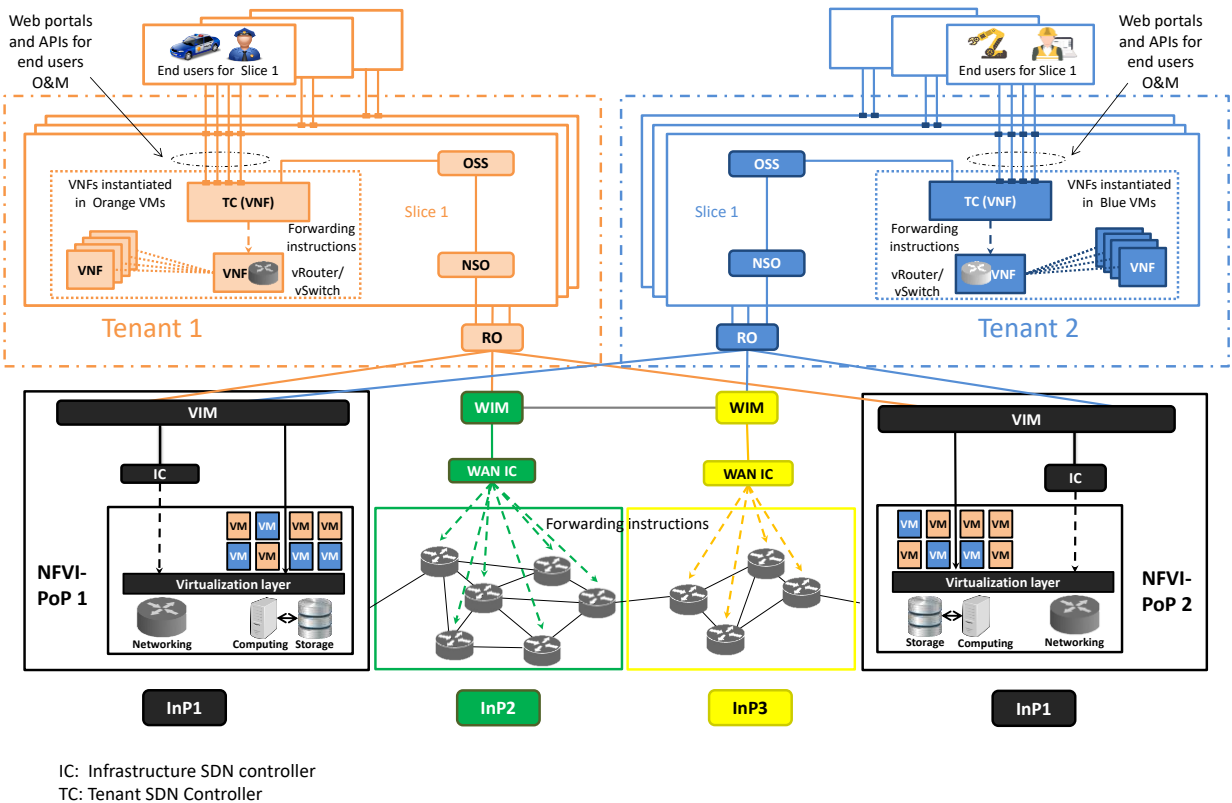

Figure 6. Network slicing deployment in a common framework, integrating both SDN and NFV.

On top of the InPs, the tenants independently manage a set of network slices. Each slice comprises an OSS, a TC, and an NSO. The OSS, an SDN application from the TC's perspective, instructs the controller to manage slice's constituent VNFs and logically compose them to efficiently realize the network service(s) the slice offers. The lifecycle of such network service(s) is managed by the NSO, which interacts with the TC via the OSS. The TC, deployed as a VNF, relies on the capabilities provided by virtual switches/routers (in the form of VNFs as well) to enable the VNF composition, forwarding pertinent instructions to such virtual switches/routers via its southbound interfaces. Through its northbound interfaces, the TC provides a mean to securely expose selected network service capabilities to end users. Such interfaces allow end users to retrieve context information (e.g. real-time performance and fault information, user policies, etc.), operate, manage and make use of the slice's network service(s), always within the limits set by the tenant.
The fact that each slice is provided with its own NSO, OSS and TC instances enables the required management isolation.

Each tenant must efficiently orchestrate their assigned resources to simultaneously satisfy the diverging requirements of the slices that are under its management. The RO is the functional block that performs such task on behalf of the tenant, providing each slice with the required resources via interfaces with each slice's NSO. The RO must perform the resource sharing among slices while fulfilling their required performance, following an adequate, effective resource management framework that must comply with both tenant and slice-specific policies. Such framework is required so that the RO enables performance isolation among slices.

All the NFVI resources available for use by a tenant (i.e. those that RO orchestrates) are supplied by the different InPs. Each InP rents part of the virtual resources according to a business lease agreement that both InP and tenant had previously signed. To access, reserve and request such resources, the tenant's RO interacts with the VIM(s) /WIM(s) by means of interfaces that those functional blocks expose and that tenant's RO consumes. Indeed, we assume that VIMs and WIMs support multi-tenancy. We also assume that WIMs can communicate with each other according to predefined business agreements. In this respect, the interaction between a WIM and an RO might be achieved indirectly through another WIM.

As Figure 6 suggests, the resource management must be performed at two levels: at the infrastructure level, where a slice-agnostic VIM/WIM provides the subscribed tenants with (virtualized) infrastructure resources, and at the tenant level, where the RO delivers its assigned resources to the corresponding slices. Both the VIM(s)/WIM(s) and the RO have to collect accurate resource usage information (each at its domain) and in turn to forecast resource availability in relatively short timescales to satisfy tenant and slice demands, respectively.

Please note that, with the exception of hardware resources, the functional blocks (e.g. VIM, RO, NSO, SDN controllers, etc.) are modeled as independent software components. The need for separate access, configuration and management suggests this modeling, wherein the software relationships are enabled with the help of the APIs that each component provides.

To preserve security and privacy isolation among slices, it is required to apply the compartmentalization principle at each virtualization level. In addition, each functional block and manageable resource (e.g. VNF) within a given slice must have its own security mechanisms, ensuring operation within expected parameters, and preventing access to unauthorized entities. This is intended to guarantee that faults or attacks occurred in one slice are confined to such slice, preventing their propagation across slice boundaries.

Additionally, although recursion has not been addressed in this example, it is readily applicable to this scenario by simply assuming some of the slice's users are tenants which in turn can deploy and operate their own slices.

# 6. Challenges and Research directions

In this section, we identify the main challenges and future research arising from implementing slicing in 5G systems.

**Performance issues in a shared infrastructure**
When network slices are deployed over a common underlying substrate, the fulfillment of performance isolation requirement is not an easy task. If tenant's RO only assigns dedicated resources to network slices, their required performance levels are always met at the cost of preventing slices to share resources. This leads to over-provisioning, an undesired situation bearing in mind that the tenant has a finite set of assigned resources. One way to resolve this issue is to permit resource sharing (see e.g. [13]), although this means slices are not yet completely decoupled in terms of performance. Thus, it is required to design adequate resource management mechanisms that enable resource sharing among slices when necessary without violating their required performance levels. To accomplish the sharing issue, the RO could use policies and strategies similar to those used in VIMs (such as the OpenStack Congress module, or Enhanced Platform Awareness attributes).

**Management and orchestration issues**
Given the dynamism and scalability that slicing brings, management and orchestration in multi-tenant scenarios are not straightforward. To flexibly assign resources on-the-fly to slices, the optimization policy that governs the RO must deal with situations where resource demands vary considerably in relatively short timescales. To accomplish this:
- An appropriate cooperation between slice-specific management functional blocks and RO is required.
- Policies need to be captured in a way that they can be automatically validated. This automation enables both the RO and slice-specific functional blocks to be authorized to perform the corresponding management and configuration actions in a timely manner.
- It is required to design computationally efficient resource allocation algorithms and conflict resolution mechanisms at each abstraction layer.

**Security and privacy**

The open interfaces that support the programmability of the network bring new potential attacks to softwarized networks. This calls for a consistent multi-level security framework composed of policies and mechanisms for software integrity, remote attestation, dynamic threat detection and mitigation, user authentication and accounting management. The security and privacy concerns arising from 5G slicing (see [14]) are today a major barrier to adopt multi-tenancy approaches.

**New business models**

The innovative partnerships between several players, each providing services at different positions of the value chain, and the integration of new tenants such as verticals, OTT service providers, and high-value enterprises, empowers promising business models. Given this business-oriented approach, new transition strategies must be broadly analyzed, allowing for a gradual evolution to future 5G networks and ensuring compatibility with past infrastructure investments. To accomplish this, a deep review of the telecom regulatory framework has to be made. Innovative ways of pricing, new grounds for cost sharing and standardized solutions, which provide the required support for interoperability in multi-vendor and multi-technology environments, must be studied as well.

# Acknowledgements

This work is partially supported by the Spanish Ministry of Economy and Competitiveness, the European Regional Development Fund (Project TIN2013-46223-P), the Ministry of Education, Culture, and Sport of the Government of Spain, under grant Beca de Colaboración (2015-2016), and the University of Granada, under grant Beca de Iniciación a la Investigación (2016-2017).

# References

[1] 5G-PPP, ERTICO, EFFRA, EUTC, NEM, CONTINUA and Networld2020 ETP, "5G empowering vertical industries", White Paper, Feb. 2016.
[2] M. Pearce, S. Zeadally, and R. Hunt, "Virtualization: Issues, security threats, and solutions", *ACM Computer Surveys (CSUR)*, vol.45, no.2, Feb. 2013, pp. 1-38.
[3] N. M. M. K. Chowdhury and R. Boutaba, "A survey of network virtualization", *Computer Networks*, vol. 54, no.5, Apr. 2010, pp. 862–876
[4] ONF TR-521, "SDN Architecture", Feb. 2016.
[5] ONF TR-526, "Applying SDN Architecture to 5G Slicing", Apr. 2016.

[6] ETSI GS NFV 002, "Network Functions Virtualization (NFV); Architectural Framework", V1.1.1, Dec. 2014.
[7] ONF TR-518, "Relationship of SDN and NFV", Oct. 2015
[8] ETSI GS NFV-EVE 005, "Network Functions Virtualisation (NFV); Ecosystem; Report on SDN Usage in NFV Architectural Framework", V1.1.1, Dec. 2015.
[9] ETSI GS NFV-MAN 001, "Network Functions Virtualisation (NFV); Management and Orchestration", V1.1.1, Dec. 2014.
[10] ETSI GS NFV-INF 001, "Network Functions Virtualisation (NFV); Infrastructure Overview", V1.1.1, Jan. 2015.
[11] ETSI GS NFV-SEC 003, "Network Functions Virtualisation (NFV); NFV Security; Security and Trust Guidance", V1.1.1, Dec. 2014.
[12] ETSI GS NFV 003, "Network Functions Virtualisation (NFV); Terminology for Main Concepts in NFV", V1.2.1, Dec. 2014.
[13] P. Andres-Maldonado, P. Ameigeiras, J. Prados-Garzon, J. Ramos-Munoz, and J. Lopez-Soler, "Virtualized MME Design for IoT Support in 5G Systems", *Sensors*, vol. 16, no. 8, Aug. 2016, pp. 1338-1362
[14] R. Harel, and S. Babbage, "5G security recommendations Package #2: Network Slicing", NGMN Alliance, Apr. 2016.

# Biographies

**Jose Ordonez-Lucena** (jordonez93@gmail.com) received his B.Sc. in Telecommunications Engineering by the University of Granada (Spain) in 2015. Currently he is a Master's student and is working in research projects with the Department of Signal Theory, Telematics and Communication of the University of Granada. His research interests are focused on network virtualization and network slicing in 5G systems.

**Pablo Ameigeiras** (pameigeiras@ugr.es) received the M.Sc.E.E. degree in 1999 from the University of Malaga, Spain. He carried his master thesis at the Chair of Communication Networks, Aachen University (Germany). In 2000 he joined the Cellular System group at the Aalborg University (Denmark) where he carried out his Ph.D. thesis. After finishing his Ph.D. he worked in Optimi/Ericsson. In 2006 he joined the University of Granada (Spain), where he has been leading several projects in the field of LTE and LTE Advanced systems. Currently his research interests include the application of the Software Defined Networking (SDN) and Network Functions Virtualization (NFV) paradigms for 5G systems.

**Diego Lopez** (diego.r.lopez@telefónica.com) joined Telefonica I+D in 2011 as a Senior Technology Expert and is currently in charge of the Technology Exploration activities within the GCTO Unit. Diego is focused on network virtualization, infrastructural services, network management, new network architectures, and network security. Diego chairs the ETSI ISG on Network Function Virtualization and the NFVRG within the IRTF, and he is member of the Board of 5TONIC, the Telefonica 5G Testbed.

**Juan J. Ramos-Munoz** (jjramos@ugr.es) received in 2001 his M.Sc. in Computer Sciences degree by the UGR (University of Granada, Spain). Since 2009, he holds a Doctorate degree from the UGR. He is a Lecturer at the Department of Signals Theory, Telematics and Communications of the UGR. He is also member of the Wireless and Multimedia Networking Lab. His research interests are focused on real-time multimedia streaming, Quality of Experience assessment, network virtualization and network slicing for 5G.

**Javier Lorca** (franciscojavier.lorcahernando@telefonica.com) received an M.Sc. degree in Telecommunication Engineering from Universidad Politécnica de Madrid (UPM) in 1998, and is currently pursuing his PhD degree in Universidad de Granada. He is in charge of Radio Access Networks Innovation within Telefónica Global CTO. His research is focused on 5G, including virtualization, massive MIMO, mmWaves, new waveforms, interference control, and advanced channel coding techniques. Javier has multiple patents and publications and a book chapter on 5G.

**Jesús Folgueira** (jesus.folgueira@telefonica.com) received his M.Sc. degree in Telecommunications Engineering from UPM (1994) and MSc in Telecommunication Economics in 2015 (UNED). He joined Telefónica I+D in 1997. He is currently the Head of Transport and IP Networks within Telefonica Global CTO unit, in charge of Network Planning and Technology. He is focused on Optical, Metro & IP Networks architecture and technology, network virtualization (SDN/NFV) and advanced switching. His expertise includes Broadband Access, R&D Management, and network deployment.